\documentclass[12pt,a4paper,fleqn]{article}
\usepackage[T1]{fontenc}
\usepackage{amsmath}
\usepackage{amssymb} 
\usepackage{bbm} 
\usepackage[small]{caption2} 
\usepackage{graphicx} 
\usepackage{mathrsfs}	
\usepackage[small,loose]{subfigure}  
\usepackage{cite} 
\usepackage{cancel}
\usepackage{psfrag} 
\usepackage{color} 
\newcommand{\eVdist}{\kern-0.06em}

\newcommand{\gev}{\:\text{Ge\eVdist V}}
\newcommand{\gv}{\:\text{G\eVdist V}}

\newcommand{\cm}{\:\text{cm}}
\newcommand{\kpc}{\:\text{kpc}}

\newcommand{\mb}{\:\text{mb}}
\newcommand{\s}{\:\text{s}}

\DeclareMathAlphabet{\mathpzc}{OT1}{pzc}{m}{it}

\allowdisplaybreaks

\begin{document}
\thispagestyle{empty}
\begin{titlepage}

\begin{flushright}
TUM-HEP-819/11
\end{flushright}

\vspace*{1.0cm}

\begin{center}
\Huge\textbf{Dark Matter after BESS-Polar II}
\end{center}
\vspace{1cm}
 \center{
\textbf{
Rolf Kappl\footnote[1]{Email: \texttt{rolf.kappl@ph.tum.de}}, 
Martin Wolfgang Winkler\footnote[2]{Email: \texttt{mwinkler@ph.tum.de}}
}
}
\\[5mm]
\begin{center}
\textit{\small
~Physik-Department T30, Technische Universit\"at M\"unchen, \\
James-Franck-Stra\ss e, 85748 Garching, Germany
}
\end{center}

\date{}
\vspace{1cm}

\begin{abstract}
The BESS-Polar collaboration has recently performed a precise measurement of the local antiproton flux which is consistent with a pure secondary production of antiprotons. We constrain a possible primary component originating from dark matter pair-annihilations. We derive limits on the annihilation cross section which are stronger than or comparable to those from the PAMELA satellite experiment for dark matter masses up to 200 GeV. Especially, we exclude thermal WIMPs with masses in the range 3-20 GeV if they annihilate dominantly into quark pairs unless their cross section is velocity suppressed.
\end{abstract}

\end{titlepage}

\newpage
\section{Introduction}

Antimatter searches provide a powerful tool to identify exotic sources of 
particles in our galaxy. In combination with other direct or indirect 
detection techniques they may eventually allow us to reveal the nature of dark 
matter. Recently, the BESS-Polar II experiment has measured the local antiproton flux in the range 
$0.2-3.5\gev$~\cite{Abe:2011nx}. This energy regime is of utmost 
importance as it could contain imprints from the evaporation of primordial 
black holes~\cite{Barrau:2001ev} or the pair-annihilation of weakly 
interacting massive particles (WIMPs) especially with masses of 
$\mathcal{O}(10\gev)$. Such light WIMPs have been proposed as a resolution 
for anomalies observed by the direct detection experiments 
DAMA~\cite{Bernabei:2008yi,Bernabei:2010mq}, CoGeNT~\cite{Aalseth:2010vx} 
and CRESST~\cite{Angloher:2011uu}. They have also been considered as the 
source of a possible gamma ray 
excess~\cite{Goodenough:2009gk,Hooper:2010mq,Hooper:2011ti} and radio 
filament structures~\cite{Linden:2011au} near the galactic center and, 
recently, also in the context of the isotropic radio 
emission~\cite{Fornengo:2011cn}. None of these hints is unambiguous: the 
direct detection data -- if interpreted in terms of dark matter -- are in 
tension with the experiments CDMS~\cite{Ahmed:2010wy} and 
XENON~\cite{Angle:2011th,Aprile:2011hi} while the indirect observations 
may find an astrophysical explanation. Nevertheless, the light WIMP 
hypothesis requires further investigation. In the same spirit, collider 
data~\cite{Goodman:2010yf,Goodman:2010ku,Fox:2011fx,Rajaraman:2011wf,Fox:2011pm}, 
the cosmic microwave background~\cite{Hutsi:2011vx,Galli:2011rz}, gamma 
rays~\cite{Cirelli:2009dv,Abazajian:2010sq,Abdo:2010dk,Arina:2010rb,Hutsi:2010ai} 
and neutrino observations at Super-Kamiokande~\cite{Kappl:2011kz} were 
used to constrain the properties of light WIMPs.

Earlier antiproton searches by BESS~\cite{Orito:1999re,Maeno:2000qx}, AMS~\cite{Aguilar:2002ad} and PAMELA~\cite{Adriani:2010rc} have been analyzed in the context of light supersymmetric dark matter~\cite{Donato:2003xg,Bottino:2005xy,Ferrer:2006hy,Bottino:2007qg,Delahaye:2011cu,Cerdeno:2011tf} and for the specific example of a 10 GeV WIMP annihilating into bottom pairs~\cite{Lavalle:2010yw}. A model-independent study, however, seems now in order, especially as the quality of data has improved substantially with the BESS-Polar II antarctic flight. But the high precision of the measurement also requires a profound understanding of the secondary antiproton background which arises from the spallation of cosmic rays on interstellar matter.

Therefore we will recalculate the secondary antiproton flux taking into account improvements in the determination of the antiproton source term and the cosmic ray propagation parameters. We will then evaluate the primary contribution from dark matter annihilation considering all relevant hadronic final states. This includes a careful treatment of processes like reacceleration and energy losses which substantially affect antiproton fluxes at low energies but have been neglected in many previous works. Finally, based on the BESS-Polar II antiproton search, we will present limits on the annihilation cross section of dark matter.

\section{Secondary Antiproton Background}

While the possibility of a primary antiproton source in our galaxy is highly speculative, the presence of so-called secondary antiprotons is well established. These originate from spallations of cosmic rays on the interstellar gas. Within error bars all antiproton searches performed so far are consistent with a purely secondary origin. Owing to the high precision of the BESS-Polar II data we decided to reevaluate the secondary antiproton flux using the 2-zone diffusion model of Donato et al.~\cite{Donato:2001ms}. Compared to the original work we improve the calculation by including a new determination of propagation parameters~\cite{Putze:2010zn} and by using the updated antiproton source term given in~\cite{Duperray:2005si}.

\subsection{Source Term}\label{sec:sourceterm}

Galactic cosmic rays which mainly consist of protons (H) and helium (He) may create antiprotons by inelastic interactions with the interstellar gas in the galactic disc. The dominant reactions are H,He$\,+\,$H$_{\text{ISM}}$,He$_{\text{ISM}}\rightarrow \bar{p} $\,+\,$ X$ where the index ISM stands for interstellar matter. The secondary source term which describes the differential $\bar{p}$ production rate per volume, time and energy takes the form
\begin{equation}\label{eq:source}
 q^\text{sec}_{\bar{p}}(T) = 2 \sum\limits_{\text{A}=\text{H},\text{He}}\;\,4 \pi\int\limits_{E_\text{th}}^{\infty} dT' \left( \frac{d\sigma}{dT}\right)_{\!\text{A}+\text{A}_{\text{ISM}}\rightarrow \bar{p}+X} n_{\text{A}_{\text{ISM}}} \;\Phi_\text{A}(T')\;.
\end{equation}
Here $\Phi_\text{A}$ denotes the cosmic ray fluxes\footnote{In principle the proton and helium fluxes change with the distance from the galactic center. We find, however, that the inclusion of the radial dependence only marginally affects the local antiproton flux. Therefore we take $\Phi_\text{H,He}$ to be homogeneous in the galactic disc.} of protons and helium which can be extracted from~\cite{Shikaze:2006je}, $n_{\text{A}_{\text{ISM}}}$ their interstellar densities ($n_{\text{H}_{\text{ISM}}} \simeq 0.9\cm^{-3}$, $n_{\text{He}_{\text{ISM}}} \simeq n_{\text{H}_{\text{ISM}}}/10$~\cite{Donato:2001ms}). The differential cross section for the reaction A+A$_{\text{ISM}}\rightarrow \bar{p}$\,+\,$ X$ is expressed in terms of the kinetic energies $T'$ and $T$ of the incoming nucleus and the outgoing antiproton respectively. The threshold energy is $E_\text{th}=6\,m_p$ and the factor of two on the right-hand side accounts for $\bar{p}$ production by anti-neutron decay.

The calculation of the source term requires a reliable parameterization of the antiproton production cross sections. The p-p interaction is well measured and an analytic expression for the cross section $\sigma_{\text{H}+\text{H}\rightarrow \bar{p}+X}$ was provided in~\cite{Tan:1982nc,Tan:1983de}. The interactions involving helium were first treated in form of a simple energy-independent enhancement factor (see e.g.~\cite{Webber:1989aa,Gaisser:1992aa}), later a model-based evaluation using the DTUNUC Monte Carlo program was performed~\cite{Simon:1998aa,Donato:2001ms}. In this work we prefer the semi-analytic approach of~\cite{Duperray:2003bd,Duperray:2005si} where cross sections are obtained by fitting the parameterization of Kalinovski et al.~\cite{Kalinovski:1989aa} to the available experimental data. We consider this method more reliable towards lower $\bar{p}$ energies where the DTUNUC model reaches the edge of its validity~\cite{Duperray:2005si}. The source term can be extracted from figure~9 in~\cite{Duperray:2005si}, a good fit in the range $T=0.1-100\gev$ is given by
\begin{equation}
 q^\text{sec}_{\bar{p}}(T) = \left(\frac{5.72\cdot10^{-30}}{\cm^{3} 
\s\gev}\right)
\times\exp\left\lbrace \sum\limits_{n=1}^5 c_n 
\,\left[ \log\left(\frac{T}{\text{GeV}}\right)\right] ^n\right\rbrace \;, 
\end{equation} with $c_1=0.98$, 
$c_2=-0.72$, $c_3=-0.021$, $c_4=0.023$ and $c_5=-0.0021$.

\subsection{Propagation}\label{sec:SecondaryPropagation}

The antiproton flux induced by the source term $q_{\bar{p}}$ is determined 
through the diffusion equation. Including all processes relevant for 
cosmic ray propagation and assuming steady state the latter can be written 
as~\cite{Berezinskii:1990aa} \begin{equation}\label{eq:diffusionequation}
 \nabla (-K \:\nabla N_{\bar{p}} + \mathbf{V}_c \,N_{\bar{p}}) + 
\partial_T (b_\text{tot} \,N_{\bar{p}} -K_{EE} \:\partial_T N_{\bar{p}} ) 
+ \Gamma_\text{ann}\,N_{\bar{p}} = q_{\bar{p}}\;, \end{equation} where 
$N_{\bar{p}}$ denotes the antiproton space-energy density. The first term 
on the left-hand side accounts for antiproton diffusion by interactions 
with inhomogeneities in the galactic magnetic field. Convection is 
included through the galactic wind velocity $\mathbf{V}_c$. The last three 
terms on the left-hand side treat energy losses, diffusive reacceleration 
and annihilations.

To simplify the diffusion equation we follow the approach of Donato et 
al.~\cite{Donato:2001ms} which we will briefly review in the 
following.\footnote{For a full numerical solution to the diffusion 
equation see~\cite{Strong:1998pw,Moskalenko:2001ya}.} The halo where 
diffusion and convection occurs is approximated by a cylinder of half 
height $L$ and radius $R$ equal to that of the galactic disc, i.e. 
$R\simeq 20\kpc$. As a boundary condition it is imposed that the 
antiproton density vanishes at the edge of the halo. The diffusion 
parameter $K$ is assumed to be homogeneous over the halo, using 
magnetohydrodynamics considerations it can be written in the 
form~\cite{Ptuskin:1997aa} \begin{equation}
 K = K_0 \,\beta \,\left(\frac{p}{\text{GeV}}\right)^\delta\;, 
\end{equation} where $K_0$ is 
a normalization constant, $\delta$ the power law index, $\beta$ and $p$ 
the antiproton velocity and momentum respectively. The galactic wind 
velocity $\mathbf{V}_c$ is taken to be constant and pointing away from the 
galactic disc.

Energy losses, reacceleration and annihilations as well as the source term are confined to the galactic disc. Neglecting the thickness of the disc one has to multiply them by $2\,h\,\delta(z)$ in order to keep a proper normalization. Here $z$ parameterizes the distance from and $h\simeq 0.1\kpc$ the half-height of the galactic disc. The term $b_\text{tot}$ includes ionization, Coulomb and adiabatic energy losses as well as reacceleration. It can be taken from~\cite{Maurin:2002ua}. The energy diffusion coefficient is also related to reacceleration, one finds~\cite{Maurin:2002hw}
\begin{equation}
K_{EE} = \frac{4}{3\,\delta \,(4-\delta^2) (4-\delta)}\,V_a^2\:\frac{\beta^2\,p^2}{K}\;,
\end{equation}
where $V_a$ is the Alfv\`en speed of magnetic shock waves in the galactic 
disc. One further has to consider the disappearance of antiprotons through annihilations with the interstellar hydrogen or helium. We use $\Gamma_\text{ann} = (n_{\text{H}_\text{ISM}} + 4^{2/3} \:n_{\text{He}_\text{ISM}})\, \sigma_\text{ann}\, \beta$ with~\cite{Protheroe:1981gj,Tan:1983de}
\begin{equation}
 \sigma_\text{ann} = \begin{cases} 661\mb \times \left(1 + 0.0115\, \left( \frac{T}{\text{GeV}}\right) ^{-0.774} - 0.948 \, \left( \frac{T}{\text{GeV}}\right) ^{0.0151}\right)\;\;\;& T < 14.6\gev\;,\\ 36\mb \times  \left( \frac{T}{\text{GeV}}\right)^{-0.5} & T\geq14.6\gev\;. \end{cases}
\end{equation}
Finally antiprotons may lose energy through inelastic (non-annihilating) scattering with the interstellar gas which leads to a redistribution of their energies. This effect can be included by introducing a tertiary source term in the galactic disc
\begin{equation}
 q^\text{ter}_{\bar{p}} (r,T)= (n_{\text{H}_\text{ISM}} + 4^{2/3} \:n_{\text{He}_\text{ISM}})
\left(\int\limits_T^{\infty} dT' \:\frac{d\sigma_{\text{non-ann}}}{d T} \,\beta'\, N_{\bar{p}} (r,T') -\sigma_{\text{non-ann}}\,\beta\, N_{\bar{p}}(r,T) \right) \;,
\end{equation}
where primed (unprimed) quantities refer to the antiprotons before (after) scattering while $r$ denotes the radial distance from the galactic center. The non-annihilating cross section can be extracted from~\cite{Tan:1983de}.
For the solution to the diffusion equation with the discussed approximations we refer to appendix~\ref{sec:solution}. The interstellar antiproton flux follows from
\begin{equation}
 \Phi^\text{IS}_{\bar{p}}=\frac{1}{4\pi}\,\beta \, N_{\bar{p}}\;. 
\end{equation} 
The solution $N_{\bar{p}}$ depends on the five free parameters $K_0$, $\delta$, $L$, $V_c$ 
and $V_a$ which can partly be fixed by observing the nuclear composition 
of cosmic rays. In~\cite{Maurin:2001sj,Maurin:2002hw,Putze:2010zn} the configurations 
which correctly reproduce the boron to carbon (B/C) ratio were determined. 

Unfortunately, the B/C ratio does not considerably constrain the size of the diffusion halo $L$. While this uncertainty only mildly affects the secondary antiproton flux it is important for a possible primary component which we will discuss later. 
The situation has slightly improved in the last years as there now exist several independent hints for $L\gtrsim4\kpc$ arising e.g.~from the observation of radioactive isotopes in cosmic rays (see 
discussion in~\cite{Lavalle:2010yw} and references therein). Further, a very recent study of radio data shows a clear preference for $L\sim 4\kpc$ while disfavoring smaller values of $L$~\cite{bringmann:2011py}. We therefore adopt 
$L=4\kpc$ in the following and take the remaining parameters from a new B/C analysis~\cite{Putze:2010zn} (NORM configuration in table~\ref{tab:proppara}). To be conservative we will later also consider a smaller $L=3\kpc$ where we 
adjust the other parameters according to figure~5 in~\cite{Putze:2010zn} 
(SMALL configuration in table~\ref{tab:proppara}).

\begin{table}[h]
\centering
\begin{tabular}{cccccc}
model&\(\delta\)&\(K_0\ (\text{kpc}^2\cdot\text{Myr}^{-1})\)
&\(L\ (\text{kpc})\)&\(V_c\ (\text{km}\cdot\text{s}^{-1})\)&\(V_a\ 
(\text{km}\cdot\text{s}^{-1})\)\\
\hline
NORM&0.86&0.0042&4&18.7&35.5\\
SMALL&0.86&0.0031&3&18.6&30.5
\end{tabular}
\caption{Propagation parameters consistent with the B/C ratio.}
\label{tab:proppara}
\end{table}

\subsection{Solar modulation}

The diffusion equation determines the interstellar antiproton flux
$\Phi_{\bar{p}}^\text{IS}$ while experiments measure the antiproton flux at the
top of the earth atmosphere $\Phi_{\bar{p}}^\text{TOA}$. The latter is
affected by solar modulation. To obtain $\Phi_{\bar{p}}^\text{TOA}$
one in principle has to solve a new transport equation which turns out
to be quite delicate as transport parameters change with time
correlated to the solar activity. Further, the magnetic field of the
heliosphere has a complicated structure and is only partly accessible
to experiments. 
We use the force-field approximation to
calculate the TOA antiproton flux~\cite{Gleeson:1968zz,Fisk:1973}
\begin{equation}
 \Phi_{\bar{p}}^\text{TOA}(T) \simeq \frac{2m_{\bar{p}}T+T^2}{ 
2m_{\bar{p}}(T+\phi)+(T+\phi)^2}\,\Phi_{\bar{p}}^\text{IS}(T+\phi)\;. 
\end{equation} In this simple approach particles and antiparticles are 
modulated in the same way and the value of the force field $\phi$ can be 
determined through observation of the proton flux. By taking the 
interstellar proton flux from~\cite{Shikaze:2006je} and extracting the TOA 
proton flux from~\cite{Sakai:2011aa}, we find $\phi=0.5\gv$. The 
BESS-Polar II data where taken in a period of lowest solar activity where the 
predictions of the force-field approximation and more sophisticated models 
of solar modulation converge. While this gives confidence that our 
treatment is sufficient, some remarks seem in order.

In general also drift effects play an essential role for the transport of 
charged particles in the heliosphere \cite{Bieber:1999dn,Potgieter:2004}. 
The latter are charge-dependent and therefore distinguish between protons 
and antiprotons. If one compares the antiproton measurements
by BESS-Polar II and PAMELA~\cite{Adriani:2010rc} one finds that the flux
determined by PAMELA is slightly higher~\cite{Abe:2011nx}. 
This difference -- if it has its origin in solar modulation 
-- can hardly be explained within the force field approximation as also 
the $\bar{p}/p$ ratio of PAMELA is larger~\cite{Sakai:2011aa}. While 
BESS-Polar II was operating in a very short time period around the solar 
minimum (December 2007 and January 2008) PAMELA took its data from 2006 to 
2008, i.e. partly at intermediate solar activity. Away from the solar 
minimum drift models like~\cite{Bieber:1999dn,Potgieter:2004} predict an 
increase of the $\bar{p}/p$ ratio which could resolve the small 
discrepancy between BESS-Polar II and PAMELA. A similar effect is also visible 
in experimental data on the $e/p$ ratio measured 
by the Ulysses spacecraft (see e.g.~\cite{Heber:2009} 
and the discussion in~\cite{Cliver:2010}).

\subsection{Comparison with Experimental Results}\label{sec:comparison}

\begin{figure}[t]
\begin{center}
  \includegraphics[width=9.5cm]{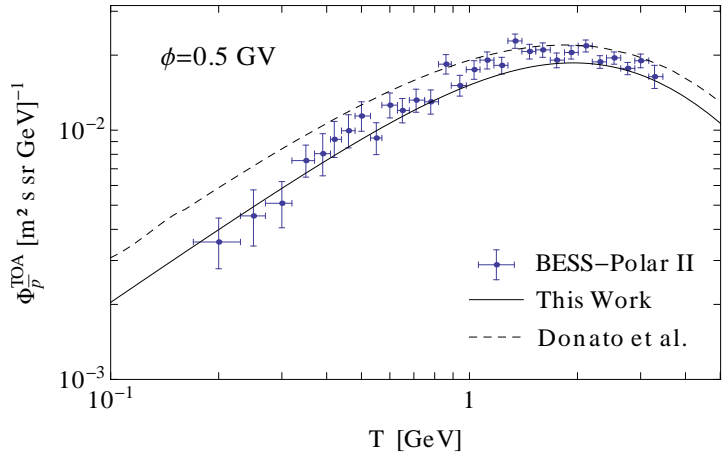}
\end{center}
\caption{TOA antiproton flux measured by BESS-Polar II together with the predicted secondary flux from our calculation and from Donato et al.. Solar modulation is included through the force field approximation with $\phi=0.5\gv$.}
\label{fig:plotsecondary}
\end{figure}

Having discussed the antiproton production and propagation we can now turn to a comparison with experimental data and earlier work. As can be seen in figure~\ref{fig:plotsecondary} the secondary antiproton flux from our calculation\footnote{The secondary antiproton fluxes for the NORM and SMALL configurations are virtually indistinguishable. Therefore, they are not discussed separately.} deviates considerably from that of Donato et al.~\cite{Donato:2001ms} although our calculation is based on the same propagation model. The reason is in part that we took the propagation parameters from the new B/C analysis~\cite{Putze:2010zn} which especially suggests a higher galactic wind velocity $V_c$. This leads to a slight decrease of the local flux as more antiprotons can escape via convection. Further, we used a different source term compared to Donato et al. which is smaller especially towards low energies (see discussion in section~\ref{sec:sourceterm}). As shown by the BESS-Polar II collaboration (figure~3 in~\cite{Abe:2011nx}) the flux of Donato et al. is to shallow to well reproduce their data at low energies even if one allows for an arbitrary normalization factor. 

If we compare our flux with the BESS-Polar II measurement we find good consistency regarding the spectral shape. On average, our flux is slightly below the measured flux. A $\chi^2$-analysis gives $\chi^2/\text{d.o.f.}=2.1$ and $\chi^2/\text{d.o.f.}=0.86$ if we would normalize our flux by a factor of $1.1$. A small underestimation of the flux can easily arise by a variety of systematic effects. We have neglected antiproton production on heavier nuclei which may increase the antiproton flux by $\sim5\%$~\cite{Simon:1998aa}. Further, the interstellar densities of H, He, the propagation parameters as well as our treatment of solar modulation are subject to uncertainties. It is thus evident that secondary antiprotons alone are capable to explain the experimental data. This was also found in~\cite{Abe:2011nx}. Nevertheless, we will take the conservative view that some fraction of the measured flux could still be of primary origin. In the following we will constrain the contribution to $\Phi_{\bar{p}}$ from dark matter annihilation.

\section{Antiprotons from Dark Matter Annihilation}

Dark matter annihilation can be an efficient source of antiprotons in our galaxy. We will now 
derive the primary antiproton flux trying to make only few assumptions on 
the nature of dark matter. As we restrict our analysis to the BESS-Polar 
II experiment we only consider the mass range $m_\chi\leq200\gev$. For 
heavier dark matter particles the PAMELA satellite experiment is more 
sensitive and we refer the reader to~\cite{Donato:2008jk}.

\subsection{Model-independent Approach}

Antiprotons can be generated if the annihilation products of dark matter particles $\chi$
involve quarks or gauge bosons. We will consider 
\begin{equation}\label{eq:SMchannels}
 \chi\chi \rightarrow u\bar{u}, \,d\bar{d}, \,s\bar{s}, 
\,c\bar{c},\,b\bar{b},\,WW,\,ZZ \; \end{equation} and assume 
$100\%$ 
annihilation into one channel. One can easily rescale our results with 
the corresponding branching fraction. Leptonic final states are omitted as 
they do not give rise to an appreciable antiproton production.\footnote{Note, however, that leptophilic annihilation may give rise to antiprotons through electroweak bremsstrahlung~\cite{Kachelriess:2009zy,Bell:2011eu,Ciafaloni:2011sa,Garny:2011cj}.}

The primary antiproton source term reads
\begin{equation}\label{eq:PrimarySource}
  q^\text{prim}_{\bar{p}}(r,z,T)  =\frac{\rho_\chi^2(r,z)}{m_\chi^2}\, 
\frac{\langle \sigma_{\text{ann}} 
v\rangle_0}{2}\,\frac{dN_{\bar{f}f}}{dT}\;. \end{equation} Here 
$dN_{\bar{f}f}/dT$ denotes the antiproton energy spectrum per annihilation 
for the channel under consideration ($f=u,\,d,\,s,\,c,\,b,\,W,\,Z$). It is determined with the PYTHIA Monte Carlo (version 
8.1)~\cite{Sjostrand:2007gs}. For the dark matter density we use a 
Navarro-Frenk-White profile \begin{equation}
 \rho_\chi(r,z) = \rho_0\, \frac{r_\odot}{\sqrt{r^2+z^2}} \left( \frac{r_c 
+ r_\odot}{r_c + \sqrt{r^2+z^2}}\right)^2 \end{equation} with 
$r_\odot=8.5\kpc$ and $r_c=24.4\kpc$. We checked that the primary antiproton 
flux is virtually insensitive to the choice of the profile. For the local 
dark matter density we take $\rho_0=0.39\gev\cm^{-3}$ 
from~\cite{Catena:2009mf}, a similar value was also found 
in~\cite{Salucci:2010qr}. Note, however, that this quantity is subject 
especially to systematic uncertainties~\cite{Pato:2010yq}.

The dark matter annihilation cross section $\langle \sigma_{\text{ann}} 
v\rangle_0$ averaged over the current velocity distribution is a free 
parameter. In case dark matter is produced thermally as suggested by the 
WIMP hypothesis it is related to the dark matter density. One has to bear 
in mind that the velocity distribution of WIMPs at freeze-out is different 
from the current one. However, in a wide class of models 
$\sigma_{\text{ann}} v$ is velocity-independent and then $\langle 
\sigma_{\text{ann}} v\rangle_0$ is fixed as~\cite{Drees:2009bi} 
\begin{equation}\label{eq:relic}
  \langle \sigma_{\text{ann}} v\rangle_0 \simeq 
10^{-26}\cm^3\s^{-1}\times\frac{1}{\sqrt{g_*(T_F)}}\:\frac{m_\chi}{T_F}\,, 
\end{equation} 
where $T_F$ is the freeze-out temperature. The effective number of degrees of freedom at freeze-out $g_*(T_F)$ can be taken from~\cite{Laine:2006cp}.

\subsection{Primary Antiproton Flux}

\begin{figure}[t]
\begin{center}
  \includegraphics[width=12cm]{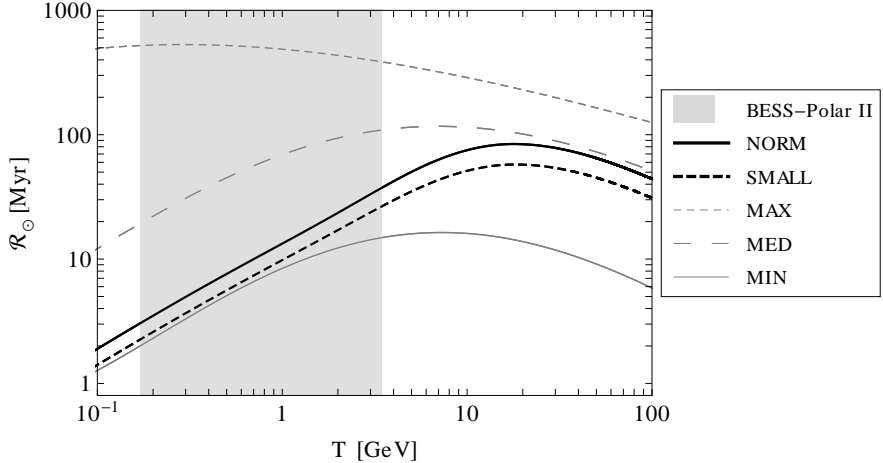}
\end{center}
\caption{Propagation function for different choices of the propagation parameters. The NORM and SMALL configurations are used in this work. They were derived from a more recent B/C analysis~\cite{Putze:2010zn} compared to MIN, MED and MAX. The shaded band refers to the energy range of BESS-Polar II.}
\label{fig:propfunction}
\end{figure}

The primary antiproton flux follows from the diffusion equation~\eqref{eq:diffusionequation} with the source term given by~\eqref{eq:PrimarySource}. The main difference compared to secondary antiprotons arises from the fact that sources are distributed over the whole diffusion halo and not just located in the galactic disc. If we neglect for the moment low energy effects on the antiprotons -- namely energy losses, reacceleration and tertiaries -- the primary antiproton space energy density can be written as (see appendix~\ref{sec:primsolution})
\begin{equation}
 N^\text{prim}_{\bar{p}} \simeq  q^\text{prim}_{\bar{p}}(r_\odot,T) \;\mathcal{R}_\odot\;,
\end{equation}
where $\mathcal{R}_\odot$ denotes the propagation function. Note that this formula only provides an estimate, in our analysis we will always use the full solution to the diffusion equation which we discuss in appendix~\ref{sec:primsolution}. Nevertheless, the function $\mathcal{R}_\odot$ nicely illustrates the dependence of $N^\text{prim}_{\bar{p}}$ on the propagation parameters. It is shown for the NORM and the SMALL configurations in figure~\ref{fig:propfunction}. To allow for comparison with earlier work we also depict $\mathcal{R}_\odot$ for the commonly used MIN, MED and MAX configurations~\cite{Donato:2003xg} which were derived from an earlier B/C analysis~\cite{Maurin:2001sj}.\footnote{The propagation function for the MIN, MED and MAX configurations was extracted from~\cite{Cirelli:2010xx}.}

At high energies cosmic ray propagation is dominated by diffusion and the differences between the five configurations mainly stem from the choice of the halo size L. The two configurations used in this paper (NORM and SMALL) approach the MED configuration as they have the same or a similar value of $L$. However, NORM and SMALL decrease more rapidly towards low energies. In this regime the process of convection rapidly gains importance, and -- as suggested by the new B/C analysis~\cite{Putze:2010zn} -- NORM and SMALL assume a larger convective wind than MIN, MED and MAX. 
In the energy range of BESS-Polar II we will therefore obtain primary antiproton fluxes in our analysis which are similar as in the MIN configuration.

\begin{figure}[t]
\begin{center}
  \includegraphics[width=7.7cm]{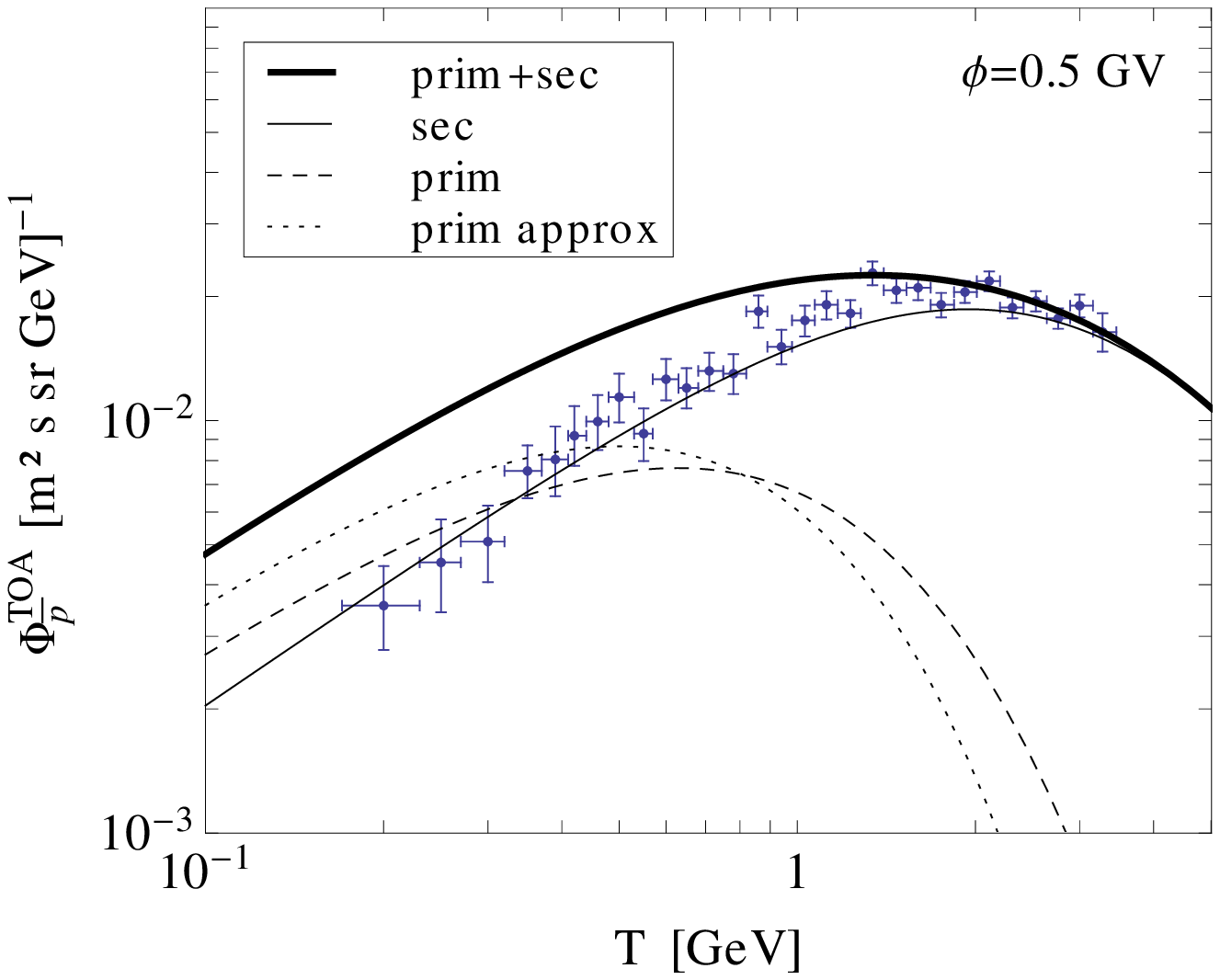}\hspace{2mm}
  \includegraphics[width=7.7cm]{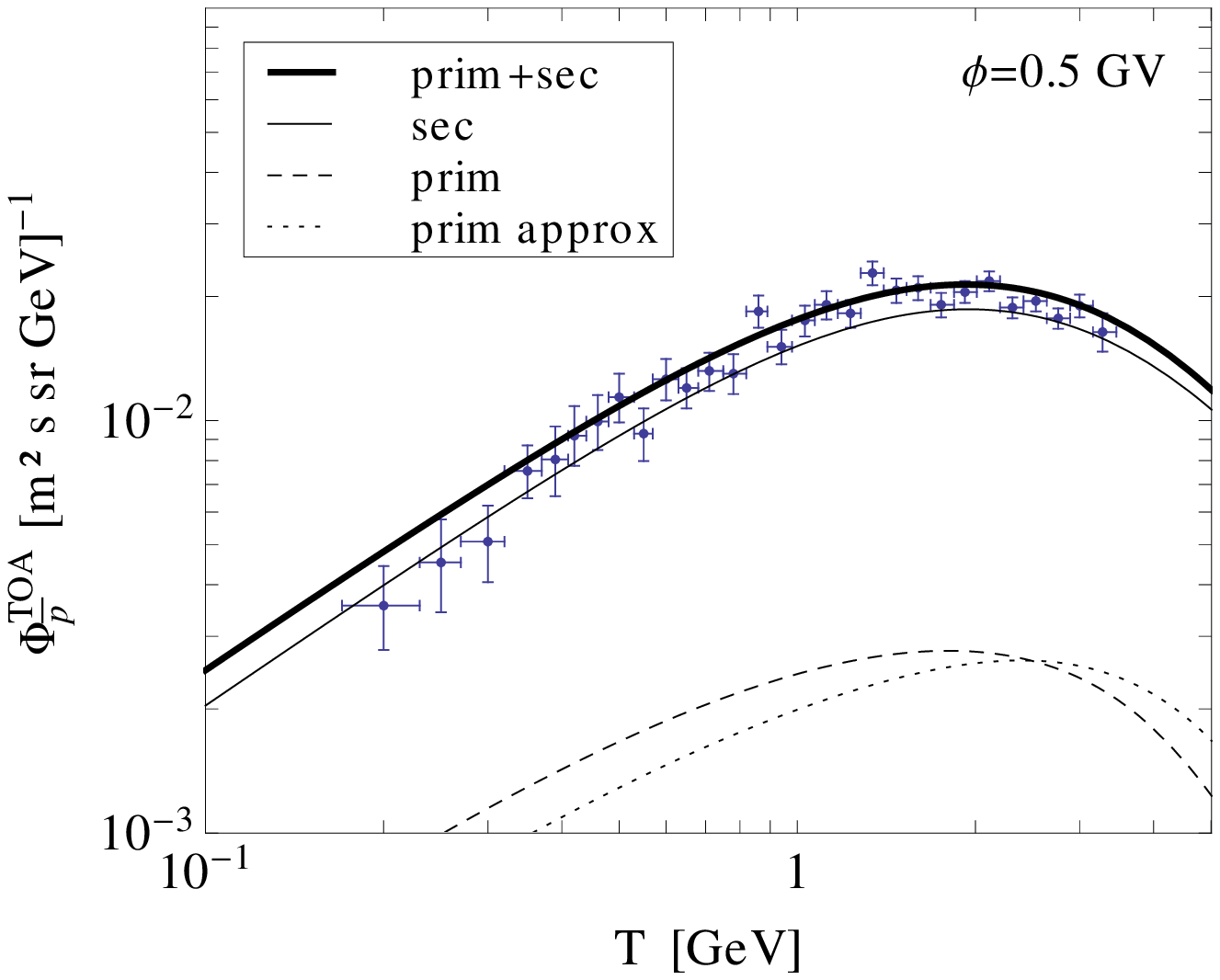}
\end{center}
\caption{Comparison of secondary (sec) and primary (prim) antiproton fluxes for the case of a dark matter particle annihilating into bottom quarks with mass $m_\chi=8\gev$ (left) and $m_\chi=30\gev$ (right). A cross section of $\langle\sigma_{\text{ann}} v\rangle_0=3\times 10^{-26}\cm^3\s^{-1}$ is assumed. Also shown are the primary fluxes without taking into account low energy effects (prim approx).}
\label{fig:primaryflux}
\end{figure}

In figure~\ref{fig:primaryflux} we depict the primary antiproton fluxes (NORM configuration) expected from a thermal WIMP annihilating into bottom quarks for two different masses. To illustrate the importance of energy losses, reacceleration and tertiaries we show the primary flux with and without taking into account these low energy effects.
For a first impression of the sensitivity of BESS-Polar II, we have also included the secondary background and the data points. It can be seen that the $8\gev$ WIMP is clearly inconsistent with the measurement, while the $30\gev$ WIMP is still viable.

\section{Limits from BESS-Polar II}

\subsection{Our Analysis}

As discussed in section~\ref{sec:comparison} the secondary antiproton flux alone is capable to explain the BESS-Polar II measurement, there is no need to include a primary component. We therefore do not try to improve the fit to the data by invoking dark matter annihilations. We rather want to determine the maximal rate of annihilations still consistent with the experiment. For this we will use a modified $\chi^2$-metric which only punishes a given configuration if the total (primary + secondary) antiproton flux overshoots the data. We define
\begin{equation}
  \chi_\text{mod}^2 = \sum\limits_i 
\begin{cases}\frac{(\overline{\Phi_{\bar{p},i}^\text{TOA}} - 
(\Phi_{\bar{p},i}^\text{data}))^2}{\sigma_{i}^2} 
&\quad \overline{\Phi_{\bar{p},i}^\text{TOA}}>\Phi_{\bar{p},i}^\text{data}\;, \\ 
0 & \quad
\overline{\Phi_{\bar{p},i}^\text{TOA}}<\Phi_{\bar{p},i}^\text{data}\;.   
\end{cases} 
\end{equation} Here $\overline{\Phi_{\bar{p},i}^\text{TOA}}$ denotes the 
predicted flux averaged over the bin $i$, $\Phi_{\bar{p},i}^\text{data}$ 
the measured flux and $\sigma_{i}$ the statistical error in the bin. Note 
that $\chi_\text{mod}^2$ does not follow an ordinary $\chi^2$ probability 
distribution. For 29 d.o.f. the 95\% upper limit on the primary flux 
corresponds to $\chi_\text{mod}^2=25.5$.

\subsection{Results}

\begin{figure}[t]
\begin{center}
  \includegraphics[width=7.7cm]{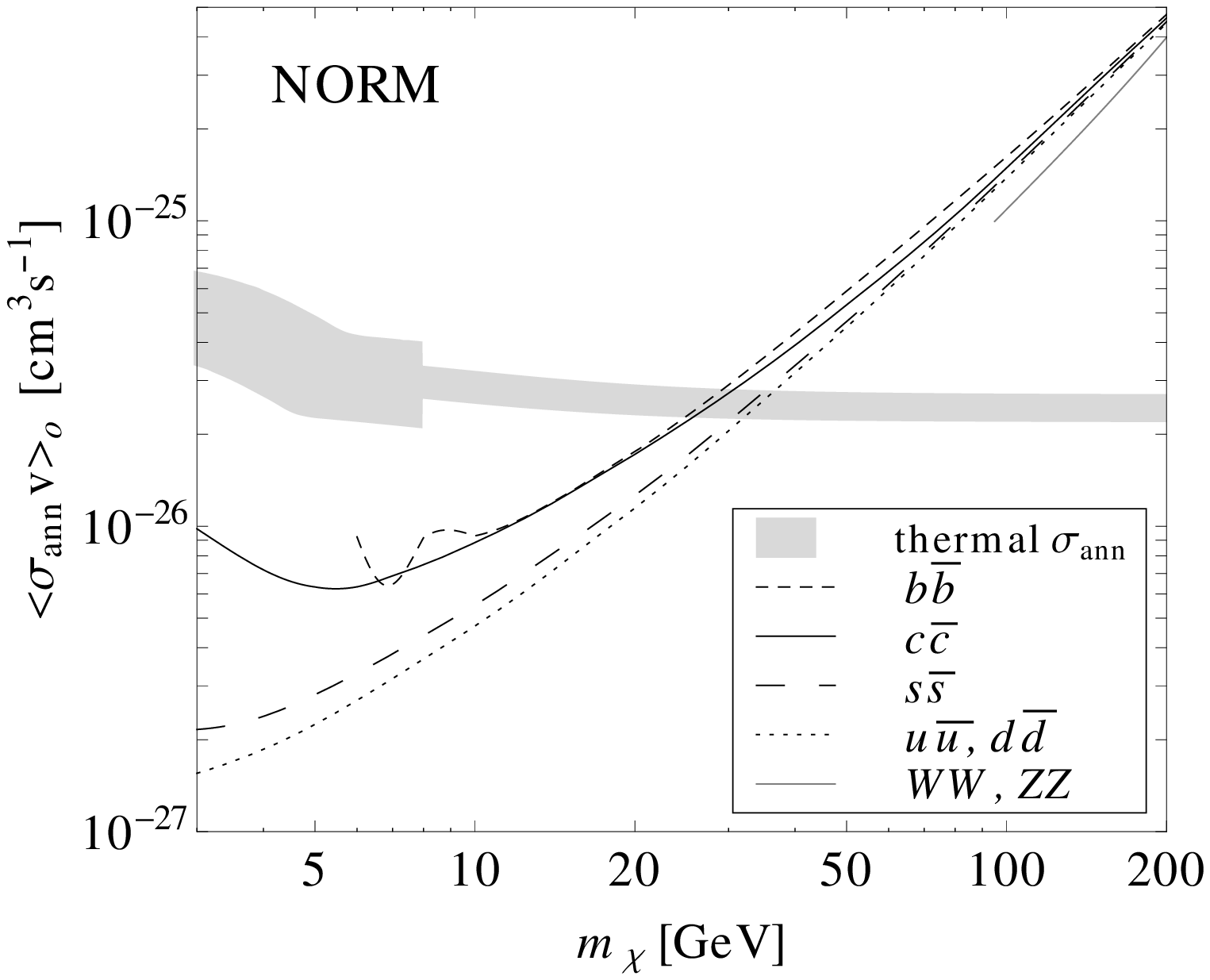}\hspace{2mm}
  \includegraphics[width=7.7cm]{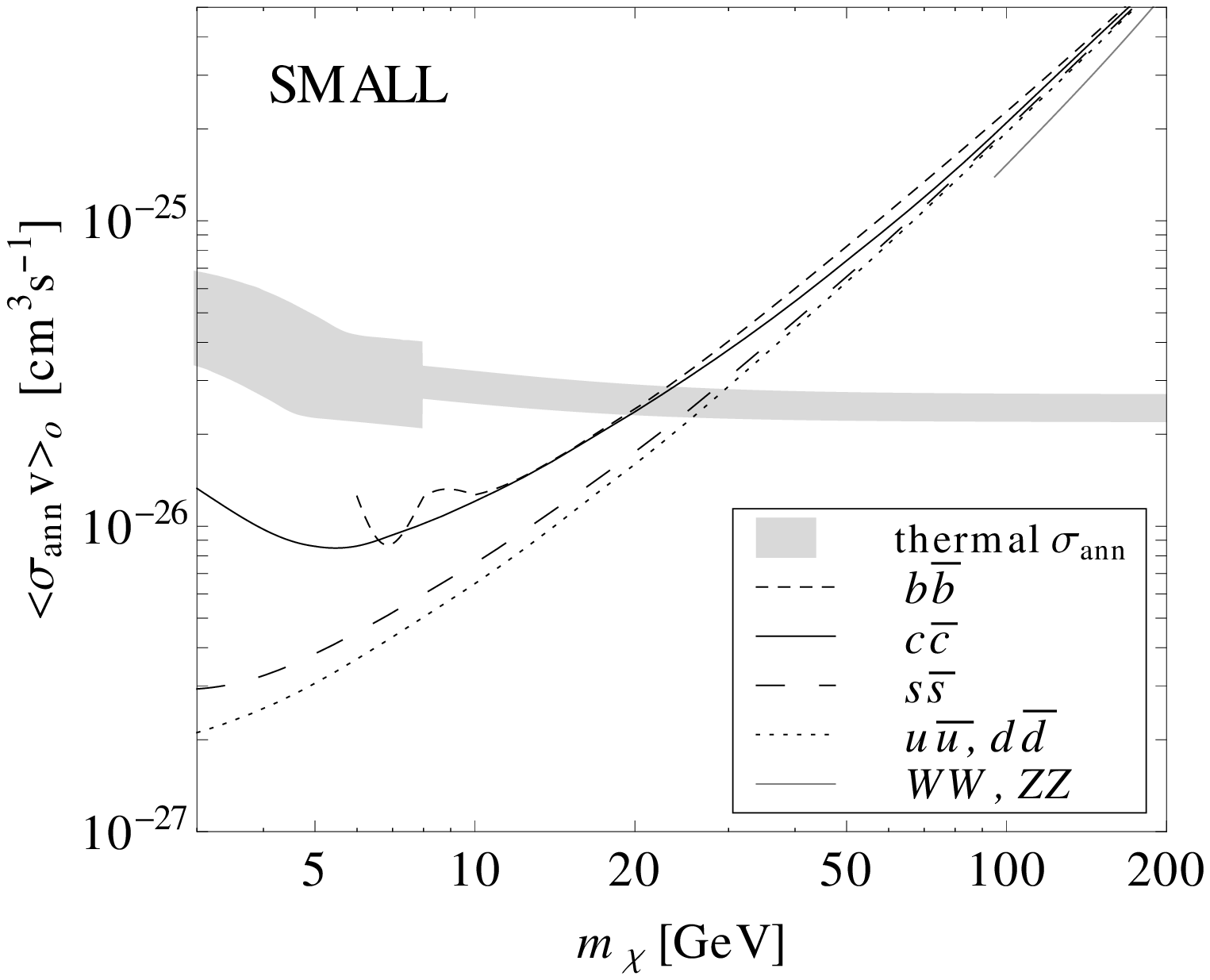}
\end{center}
\caption{Limits on the dark matter annihilation cross section for the NORM and SMALL configurations. The shaded band refers to the annihilation cross section of a thermal WIMP in the case of s-wave annihilation (see text).}
\label{fig:limits}
\end{figure}

In figure~\ref{fig:limits} we provide 95\% upper limits on the dark matter annihilation cross section for all hadronic channels. We show the results separately for the two propagation parameter sets of table~\ref{tab:proppara}. The NORM configuration assumes a diffusion halo size $L=4\kpc$ compared to $L=3\kpc$ in the SMALL configuration, the latter resulting in more conservative constraints. We also depict the annihilation cross section of a thermal WIMP for the case of s-wave annihilation. The latter is obtained from~\eqref{eq:relic} where we take $m_\chi/T_F$ to be in the range $20-25$ and include an additional $20\%$ uncertainty on the cross section if the freeze-out happens around the QCD phase transition.

It can be seen that a thermal WIMP with s-wave annihilation into quarks is excluded in the mass range $3-20\gev$ even for the SMALL configuration. At smaller masses the antiproton production threshold is approached and predictions become uncertain. The limits for the bottom and charm channels are weaker than those for light quarks especially towards low WIMP masses. The reason is that the heavy quarks induce fewer antiprotons and the spectrum is more compressed towards lower energies, i.e. more antiprotons reside below the energy threshold.

Constraints also arise for heavier dark matter particles. The primary antiproton flux from a $100\gev$ WIMP annihilating into W pairs is e.g. still to $\sim 30\%$ in the energy range of BESS-Polar II ($0.2-3.5\gev$). Nevertheless, the limits become weaker as the number density of dark matter scales inversely with its mass. Therefore, above $30\gev$ only scenarios with a non-thermal dark matter annihilation cross section can be constrained. 

\section{Conclusion}

We have studied the production and propagation of antiprotons in the light of new experimental data by BESS-Polar II. After having recalculated the cosmic ray induced flux we confirmed that the measurement is compatible with a purely secondary origin of antiprotons. We determined the primary flux which would arise from dark matter annihilations considering all relevant annihilation channels and taking into account updates on the propagation parameters. The latter especially show a trend towards higher convective winds which considerably reduces the primary antiproton flux. Nevertheless, because of the high precision of the BESS-Polar II data, we were able to provide strong constraints on dark matter annihilations. We found that a thermal WIMP with $m_\chi = 3-20\gev$ is excluded if it dominantly annihilates into quarks unless the cross section is velocity suppressed. While this statement relies on rather conservative astro- and nuclear physics assumptions it is impossible to exclude all sources of uncertainty. In this light it is important to note that our result is confirmed independently by observations of gamma rays from dwarf galaxies~\cite{Abdo:2010dk} and of the cosmic microwave background~\cite{Hutsi:2011vx,Galli:2011rz}. As the uncertainties affecting the three analyses are widely uncorrelated the exclusion of light WIMPs with hadronic annihilations seems relatively robust.

This implies that the signals of the direct detection experiments DAMA, CoGeNT and CRESST -- if interpreted in terms of dark matter -- require non-hadronic channels, a non-thermal cross section or p-wave annihilation. The first of the named options is again subject to strong constraints especially from the Super-Kamiokande neutrino telescope~\cite{Kappl:2011kz}.

Although the BESS-Polar II experiment took its data at low energies it can still provide important limits on dark matter particles with masses up to $\mathcal{O}(100\gev)$ as antiprotons typically only carry a small fraction of the energy of the mother particle. Thermal WIMPs with $m_\chi > 30\gev$ are still viable, but certain models which involve boost factors for dark matter annihilation or non-thermal cross sections can be constrained. 

\appendix

\section{Solving the Diffusion Equation}

\subsection{Secondaries}\label{sec:solution}

To get rid of the radial part of the diffusion equation one can Bessel expand the space-energy density and the source terms
\begin{equation}\label{eq:besselexpansion}
 N_{\bar{p}} (r,T) = \sum\limits_{i=1}^\infty N_{\bar{p},i}(T) J_0\left(\zeta_i\frac{r}{R}\right)\;,\quad
 q_{\bar{p}} (r,T) = \sum\limits_{i=1}^\infty q_{\bar{p},i}(T) J_0\left(\zeta_i\frac{r}{R}\right)\;.
\end{equation}
Here $R\simeq20\kpc$ denotes the radius of the galactic disc and $\zeta_i$ the $i$-th zero of the Bessel function $J_0$. The Bessel coefficients for the secondary source term are determined by 
\begin{equation}\label{eq:besselq}
 q^\text{sec}_{\bar{p},i}(T) = \frac{2}{J_1^2(\zeta_i)\,R^2} \int\limits_0^R dr \,r\, q_{\bar{p}}(r,T) J_0\left(\zeta_i\frac{r}{R}\right)\;.
\end{equation}
As discussed in section~\ref{sec:sourceterm} we assume that $q^\text{sec}_{\bar{p}} (r,T) = q^\text{sec}_{\bar{p}}(T) \; \theta(R-r)$ where $\theta$ denotes the Heaviside function and $q^\text{sec}_{\bar{p}}(T)$ is given by~\eqref{eq:source}.

Plugging~\eqref{eq:besselexpansion} into~\eqref{eq:diffusionequation} and performing the approximations described in section~\ref{sec:SecondaryPropagation} the axial part of the diffusion equation can be solved analytically. One arrives at a differential equation in energy which reads (at $z=0$)
\begin{equation}\label{eq:diffeq}
 A_{\bar{p},i} (N_{\bar{p},i}-N_{\bar{p},i}^0) +  2 h \partial_T (b_\text{tot} \,N_{\bar{p},i} -K_{EE} \:\partial_T N_{\bar{p},i} ) =  2 h q^\text{ter}_{\bar{p},i}\;
\end{equation}
with
\begin{equation}
 A_{\bar{p},i} = 2 h \Gamma_\text{ann} + V_c + K S_i \coth{\left( \frac{S_i L}{2}\right) }\;,\qquad
S_i = \sqrt{\frac{V_c^2}{K^2}+4\,\frac{\zeta_i^2}{R^2}}\;.
\end{equation}
and
\begin{equation}
 N_{\bar{p},i}^0 = \frac{ 2 \,h \,q^\text{sec}_{\bar{p},i}}{A_{\bar{p},i}}\;.
\end{equation}
The equation for the antiproton space energy density has to be solved numerically. In the high energy regime ($T\gg10\gev$), energy losses, reacceleration as well as tertiaries can be neglected and $N_{\bar{p},i}$ approaches $N_{\bar{p},i}^0$.

\subsection{Primaries}\label{sec:primsolution}

For the case of primary antiprotons, the diffusion equation can be solved in the same way as for secondaries, the only difference being the distribution of sources. It turns out that the differential equation~\eqref{eq:diffeq} is still valid, however, $N_{\bar{p},i}^0$ has to be defined differently. One finds (see e.g.~\cite{Donato:2003xg}) 
\begin{equation}
 N_{\bar{p},i}^0 =   q^\text{prim}_{\bar{p}}(r_\odot,T) \; \mathcal{R}_i
\end{equation}
with
\begin{equation}
 \mathcal{R}_i = \frac{1}{\rho_0^2} \;\frac{2\int\limits_0^L dz' \exp\left(\frac{V_c (L-z')}{2 K}\right) \sinh\left(\frac{S_i(L-z')}{2}\right)(\rho_\chi^2)_i}{A_{\bar{p},i} \sinh\left(\frac{S_i L}{2}\right) } \;\exp\left(\frac{-V_c L}{2 K}\right)\;.
\end{equation}
Here we have Bessel expanded the dark matter density distribution $\rho_\chi^2(r,z)$ and denoted the Bessel coefficients by $(\rho_\chi^2)_i$. The latter are determined analogous to the $q^\text{sec}_{\bar{p},i}$ in~\eqref{eq:besselq}.

Again the equation for $N_{\bar{p},i}$ has to be solved numerically. In the high energy limit $N_{\bar{p},i}$ approaches $N_{\bar{p},i}^0$ and the local antiproton space-energy density reads
\begin{equation}\label{eq:appr}
 N_{\bar{p}} \simeq  q^\text{prim}_{\bar{p}}(r_\odot,T) \;\mathcal{R}_\odot\;, \qquad \mathcal{R}_\odot=\sum_{i=1}^\infty \mathcal{R}_i J_0\left(\zeta_i\frac{r_\odot}{R}\right)\;.
\end{equation}
Note, however, that for antiproton energies $T \lesssim 10\gev$ this equation should only be used as a first estimate. In our analysis we will consider the full numerical solution to the diffusion equation and only use~\eqref{eq:appr} for illustrative purposes.

\section*{Acknowledgments}

We would like to thank Stefan Vogl and David Tran for helpful discussions.
This research was supported by the DFG
cluster of excellence Origin and Structure of the Universe, and the
SFB-Transregio 27 "Neutrinos and Beyond" by Deutsche
Forschungsgemeinschaft (DFG).

\bibliography{antiproton}
\bibliographystyle{ArXiv2}
\end{document}